# Boron-related defects in n-type 4*H*-SiC Schottky barrier diodes

Tihomir Knežević[1], Eva Jelavić[2], Yuichi Yamazaki[3], Takeshi Ohshima[3], Takahiro Makino[3], and Ivana Capan[1]*

[1]*Ruđer Bošković Institute, Bijenička 54, 10 000 Zagreb, Croatia*
[2] *Faculty of Science, University of Zagreb, Bijenička 32, 10 000 Zagreb, Croatia*
[3]*National Institutes for Quantum Science and Technology, 1233 Watanuki, Takasaki*

**Abstract:** We report on boron-related defects in the low-doped n-type (nitrogen-doped) 4*H*-SiC semitransparent Schottky barrier diodes (SBDs) studied by minority carrier transient spectroscopy (MCTS). An unknown concentration of boron was introduced during chemical vapor deposition (CVD) crystal growth. Boron incorporation was found to lead to the appearance of at least two boron-related deep-level defects, namely, shallow (B) and deep boron (D-center), with concentrations as high as $2 \times 10^{15}$ cm$^{-3}$. Even though the boron concentration exceeded the nitrogen doping concentration by almost an order of magnitude, the steady-state electrical characteristics of the n-type 4*H*-SiC SBDs did not deteriorate.

The earliest experimental studies of boron-related deep-level defects in SiC using deep-level transient spectroscopy (DLTS) date back to the 1980s. The pioneering work was made by Anikin et al.[1] on boron-doped 6*H*-SiC p$^+$n diodes. They not only reported the peak labeled as the D-center but also indicated that the D-center consists of two overlapping peaks with activation energies for hole emission of $E_v + 0.63$ eV and $E_v + 0.73$ eV. This result was recently confirmed experimentally using the Laplace minority carrier transient spectroscopy (Laplace-MCTS).[2] Capan et al.[2] provided direct evidence that the D-center consists of two components (i.e. emission lines), and labeled these as D1 and D2, respectively. The estimated activation energies for hole emission are $E_v + 0.49$ eV and $E_v + 0.57$ eV. They are assigned to the deep boron (isolated boron sitting at the C site) (-*h* and -*k* site, respectively). The intensity ratio D1:D2 is approximately 1:1, indicating that the B$_C$(*h*) and B$_C$(*k*) sites are equally occupied.

Suttrop et al.[3] have made significant progress in understanding boron-related defects, as it became evident that boron introduces two electrically active deep-level defects into SiC material. In their study, they used 6*H*-SiC grown by a liquid phase epitaxy (LPE) process. Boron was introduced either by ion implantation or during the LPE process from a B-doped silicon melt, and this time two electrically active deep-level defects were detected. The shallow boron at $E_v + 0.30$ eV, and, already known, deep boron (D-center) at $E_v + 0.58$ eV.

Together with aluminum, boron is the most common p-type dopant in SiC. However, unintentional incorporation of boron can occur during crystal growth. This incorporation has been previously reported and explained by the presence of boron in the graphite susceptor used for CVD growth. [4-6] The concentration of unintentionally incorporated boron can reach more than $10^{13}$ cm$^{-3}$ yielding the appearance of boron-related deep levels in the lower part of the bandgap.[2,6] This problem was studied by Storasta et al.[7] They applied MCTS to check the influence of unintentional boron incorporation during growth in low-doped n-type 4*H*-SiC. The SiC material was grown by hot-wall chemical vapor deposition (CVD). Their results also confirmed previous findings that boron introduces two deep levels, shallow (B) and deep boron (D-center) with activation energies of $E_v + 0.27$ and $E_v + 0.67$ eV, respectively. Moreover, they concluded that residual boron strongly affects the properties of SiC material grown by the CVD method. Boron-related defects act as hole traps, have large capture-cross sections, and significantly reduce the minority carrier lifetime.

In this work, we have performed electrical characterization employing temperature-dependent current-voltage (*I-V*), capacitance-voltage (*C-V*), DLTS, and MCTS measurements to verify the effect of unintentionally incorporated boron on the steady-state electrical performance of n-type 4*H*-SiC semitransparent Schottky barrier diodes. One of the main goals of this work is to show how MCTS can be utilized in the study of boron-related deep-level defects in n-type 4*H*-SiC materials.

Schottky barrier diodes (SBDs) were fabricated on nitrogen-doped (~3 × 10$^{14}$ cm$^{-3}$) 4*H*-SiC epitaxial layers with a thickness of approximately 25 µm. The n-type epi-layer was grown on an 8° off-cut silicon face of a 350 µm thick 4H-SiC (0001) wafer without a buffer layer by CVD. [8] Semitransparent SBDs for MCTS measurements were formed by the evaporation of a thin film of nickel (15 nm) through a metal mask with openings of 2 mm × 2 mm. For wire bonding, a thick nickel film (100 nm) was stacked on one corner of the Ni thin film. The same semitransparent diodes were used for all experimental techniques. No high-temperature annealing was performed before electrical characterization.

The quality of the fabricated semitransparent SBDs was evaluated by temperature-dependent *I–V* and *C–V* measurements using a Keithley 4200 SCS (Keithley Instruments, Cleveland, OH, USA). Electrically active defects were characterized by DLTS (electron traps) and MCTS (hole traps). DLTS measurements were performed in the temperature range of 100 to 450 K. The temperature ramp rate was 2 K/min. Capacitance transients were measured using a Boonton (Boonton Electronics, Parsippany, NJ, USA) 7200 capacitance meter using a 30 mV, 1 MHz sinusoidal signal. Voltage settings were reverse bias voltage, $V_R$ = -10 V, pulse bias, $V_P$ = -0.1 V. The pulse duration, $t_P$, was 10 ms. MCTS measurements were carried out using an experimental setup consisting of a Boonton 7200 capacitance meter and NI PCI-6251 DAQ. For optical excitation, a 365 nm LED (ThorLabs M365D2 LED) with a Thorlabs LDC205C LED driver was used. For the depth-profiling measurements, the reverse bias voltage was varied from -10 V to -1 V.

The steady-state electrical properties of the fabricated n-type 4*H*-SiC SBDs were checked by temperature-dependent *I-V* and *C-V* measurements. Figures 1a and 1b show respectively the *I-V* and *C-V* characteristics at temperatures from 100 K to 450 K. The ideality factor at 300 K was extracted to a value of ~1.01. The forward *I-V* characteristics at temperatures below 200 K revealed two distinct regions separated by a kink that is attributed to the presence of a Schottky barrier inhomogeneity affected either by surface defects or doping inhomogeneities.[9-11] At higher temperatures, only the dominant Schottky barrier determined the saturation current, which remained below 50 pA even for a reverse bias voltage of 100 V and at the operating temperature of 450 K. The low leakage current of 4H-SiC SBD indicates the absence of generation-recombination mechanisms, even for a fully depleted epitaxial layer, that could otherwise degrade the electrical performance. Apart from the effect of temperature on series resistance, the *C-V* characteristics measured at 1 MHz and temperatures from 100 K to 450 K show no impact of additional charge carriers on the capacitance.

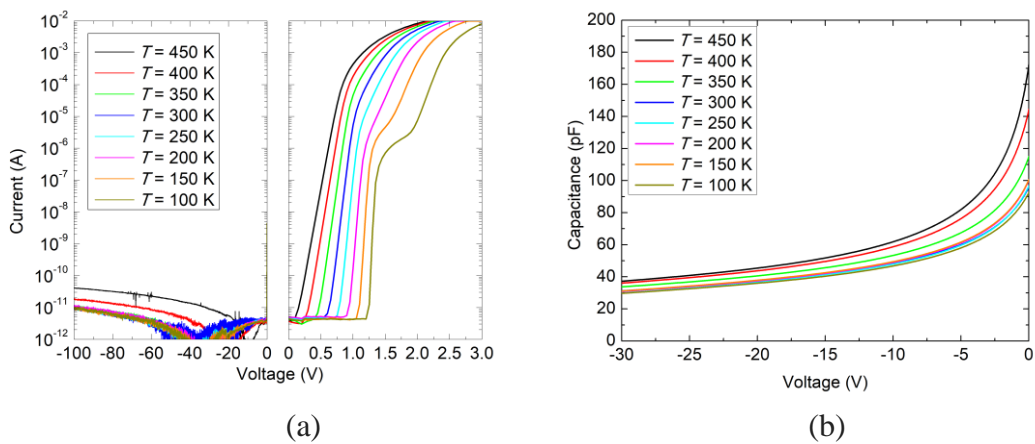

(a)          (b)

Figure 1. (a) *I-V* and (b) *C-V* characteristics for as-grown n-type 4H-SiC SBD measured at temperatures from 100 K to 450 K.

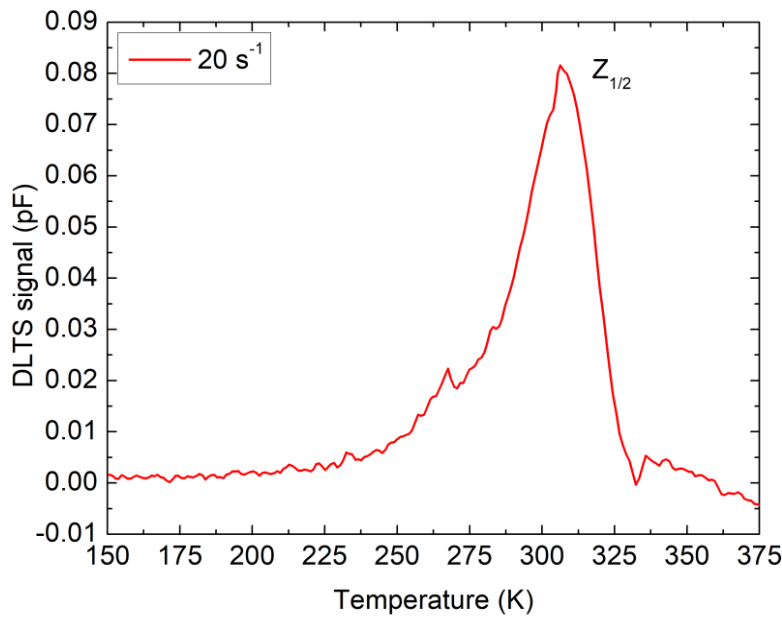

Figure 2. DLTS spectrum for as-grown n-type 4H-SiC SBD. Measurement settings were $V_R$ = -10 V, $V_P$ = -0.1 V, and $t_P$ = 10 ms.

Figure 2 shows the DLTS spectrum for the as-grown n-type 4*H*-SiC SBD. Only one, slightly asymmetric, peak with a temperature maximum of around 310 K is observed. The peak, labeled as $Z_{1/2}$, was previously assigned to a transition between the double negative and neutral charge states of carbon vacancy $V_C$ (=/0).[12] Additionally, this asymmetric DLTS peak was resolved into two components, $Z_1$(=/0) and $Z_2$(=/0), using the high-resolution Laplace DLTS measurements, which were assigned to carbon vacancies residing on two different lattice sites with local cubic (-*k*) and hexagonal (-*h*) symmetry, respectively.[13,14] Carbon vacancy is often labeled as a "lifetime killer" due to its detrimental effect on carrier lifetime[15] and it is one of the most studied and dominant defects in 4*H*-SiC material. The activation energy for electron emission was estimated from the Arrhenius plot as $E_c$ - 0.65 eV. We also estimated the $Z_{1/2}$ concentration to be $3 \times 10^{12}$ cm$^{-3}$. The obtained value is compatible with all previously published results for the low-doped n-type 4*H*-SiC SBDs.[13,14,16,17]

Since the unknown boron concentration was incorporated into the SiC material during CVD growth, there was a need for an experimental technique that will enable us to study the lower part of the bandgap, i.e., to study the minority carrier traps. It is worth noting that minority carrier traps are studied to a smaller extent compared to majority carrier traps in SiC, even though they could play an even more notable role as "lifetime-killers" than the widely known $V_C$.[7,18]

In this work, we carried out MCTS measurements to investigate the presence of boron in n-type 4*H*-SiC material. Figure 3 shows the MCTS spectrum for as-grown material. The boron concentration was high enough to introduce two known boron-related deep-level defects, labeled as the B and D-center. Together with the B and D-center, two smaller peaks, labeled X and Y, are observed in the MCTS spectrum. The activation energy for hole emission for these defects could not be accurately calculated, but comparison shows that the X resembles the defect recently reported by Fur et al.[19] They estimated the activation energy for X as $E_V$ + 0.195 eV. For Y, the activation energy as well as the origin are still unclear. Further studies are needed to find out if X and Y are boron-related defects or not.

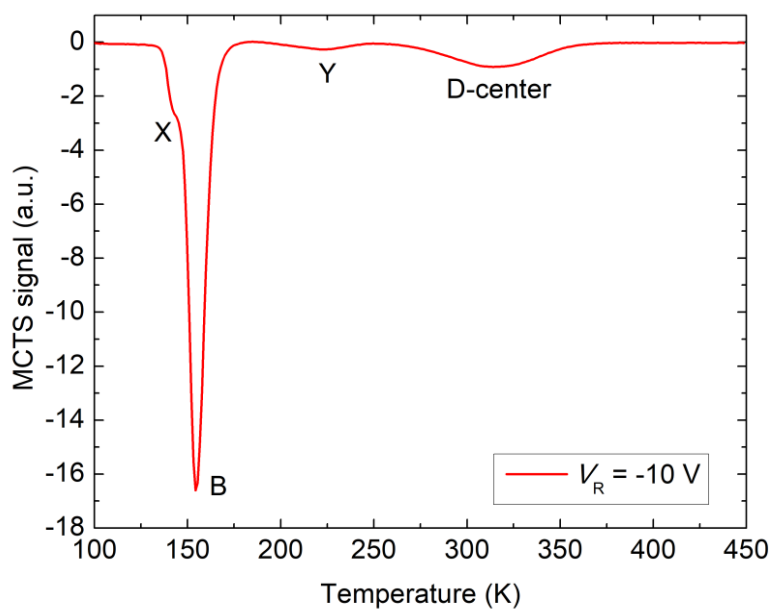

Figure 3. MCTS spectrum for as-grown n-type 4H-SiC SBD. Measurement settings were $V_R$ = -10 V, and the optical pulse width 10 ms.

From the Arrhenius plots, the activation energies for hole emissions for B and D-center are estimated as $E_V$ + 0.21 and $E_V$ + 0.60 eV, respectively. They are well-known defects and have already been assigned to shallow boron (substitutional boron at the silicon site, $B_{Si}$) and deep boron (substitutional boron at the carbon site, $B_C$). [2,7,18,20]

The intensity of the B peak is striking (Figure 3), and it is evident that the shallow boron concentration (B) is not only significantly higher compared to deep boron (D-center), but to $Z_{1/2}$ (Figure 2) as well. Since shallow boron occupies Si-sites, and deep boron occupies C-sites in the 4H-SiC materials, our results show that SiC grew under C-rich growth conditions since many empty Si-sites were available for boron.[21]

We conducted depth profiling measurements using the C-V and MCTS to acquire additional information on boron incorporation and distribution. Figure 4 shows concentration profiles as a function of depth for B and D-center, as well as the free carrier concentration profile.

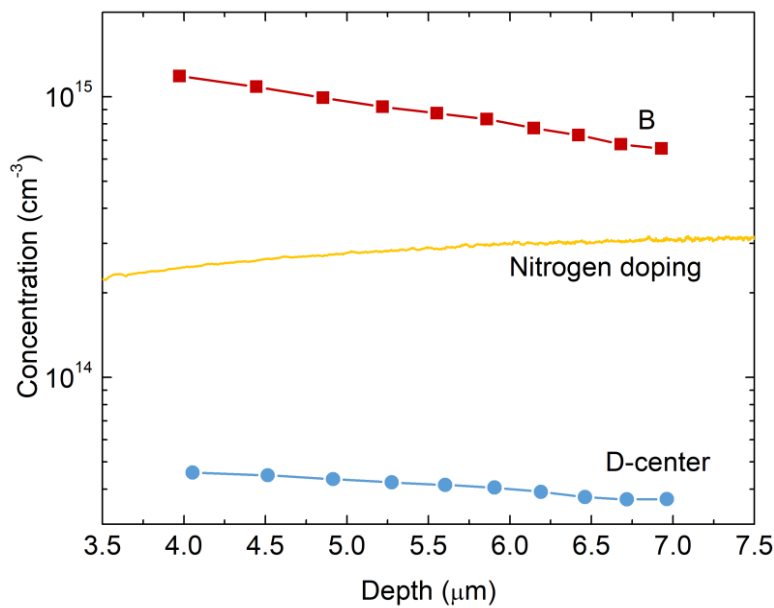

Figure 4. Depth profiles for B and D-center in n-type 4H-SiC SBD measured by MCTS. The free carrier concentration (nitrogen doping) profile was estimated from C-V measurements at room temperature (RT).

The concentration of both, B and D-center is increasing towards the surface. Moreover, the concentration of shallow boron, B, is more than an order of magnitude higher than the concentration of deep boron, D-center. This is consonant with the previously mentioned C-rich CVD growth conditions where boron easily finds empty Si-sites near the surface.

Yang et al.[6] studied the correlation between the introduced boron concentration and the concentration of the D-center using MCTS and secondary-ion mass spectrometry (SIMS) measurements. They found that the ratio between the D-center and the boron concentration is 0.02. If we take this ratio and the estimated concentration of the D-center from Figure 4 as 4.5

× $10^{13}$ cm$^{-3}$, we obtain 2.25 × $10^{15}$ cm$^{-3}$ as the boron concentration in the SiC material. This is almost an order of magnitude higher than the nitrogen doping concentration (Figure 4), and an order of magnitude higher than the concentration of unintentionally introduced boron during CVD growth, as previously assumed. [2,4,6] It is noteworthy that such a high boron concentration did not affect the steady-state electrical properties of the n-type 4*H*-SiC SBDs, as we show in Figure 1.

Another interesting feature is boron diffusion in SiC, which is extremely important from a technological aspect. Most of the available studies agree that the fast diffusion of shallow baron ($B_{Si}$) is explained by two mechanisms, the interstitial-mediated and the kick-out mechanisms[22]:

$$B_{Si} \leftrightarrow B_i + V_{Si}$$
$$B_{Si} + Si_i \leftrightarrow B_i,$$

where $B_{Si}$, $B_i$, $V_{Si}$, and $S_i$ denote substitutional boron at the Si-site, interstitial boron, silicon vacancy, and silicon interstitial, respectively.

Regarding the deep boron ($B_C$), Bockstedte et al.[23] described the preferential formation of deep boron in diffusion tails and suggested that the appearance of deep boron is more favorable compared to shallow boron under nonequilibrium conditions.

From Figure 4, we can also see a slight difference in the slope of measured depth profiles. While the concentration of D-center is more uniformly distributed over the measured volume (4 ×$10^{13}$ cm$^{-3}$ < [D] < 5 ×$10^{13}$ cm$^{-3}$), the concentration of B changes more rapidly (1.2 × $10^{15}$ cm$^{-3}$ < [B] < 6 ×$10^{14}$ cm$^{-3}$). The examined volume is rather limited, and the changes in concentrations are not too significant, therefore we cannot draw definite conclusions. For accurate and complete diffusion analysis, high-temperature (1700 - 2100 °C) annealing and SIMS measurements are a prerequisite. However, the observed dependencies in the depth profiles could be tentatively explained by different diffusion mechanisms for shallow boron ($B_{Si}$) and deep boron ($B_C$), as described above.

MCTS measurements provided direct evidence that unintentional boron incorporation during CVD growth has resulted in the appearance of at least two boron-related deep-level defects, shallow boron (B) and deep boron (D-center). These defects have already been assigned to substitutional boron occupying the Si-site ($B_{Si}$, shallow boron), and the C-site ($B_C$, deep boron). The estimated boron concentration (~2 × $10^{15}$ cm$^{-3}$) has exceeded the nitrogen doping concentration (~3 × $10^{14}$ cm$^{-3}$), however, the steady-state electrical performance of the n-type 4*H*-SiC SBDs has been preserved.

**ACKNOWLEDGEMENT**
This work was supported by the North Atlantic Treaty Organization Science for Peace and Security Program through Project No. G5674.